\documentclass[conference]{IEEEtran}
\IEEEoverridecommandlockouts
 \usepackage{lipsum,graphicx,multicol}
\usepackage{color,soul}
\usepackage{mathtools}
\usepackage{graphicx}
\usepackage{color,soul}
\usepackage{setspace}
\usepackage{multirow}
\usepackage{tablefootnote}
\usepackage[norelsize, linesnumbered, ruled, lined, boxed, commentsnumbered]{algorithm2e}
\usepackage[normalem]{ulem}
\usepackage[font=small,labelfont=bf]{caption}
\usepackage{subcaption}
\usepackage{slashbox}
\usepackage{bm, amsmath, amssymb, amsthm}
\usepackage{amsfonts}
\usepackage {amssymb,amsmath,xcolor}
\usepackage{booktabs}
\usepackage{cite}
\usepackage{array}
\usepackage[acronym]{glossaries}
\usepackage{cases}
\usepackage{cleveref}

\newtheorem{Proposition}{Proposition}

\def\BibTeX{{\rm B\kern-.05em{\sc i\kern-.025em b}\kern-.08em
    T\kern-.1667em\lower.7ex\hbox{E}\kern-.125emX}}
    
  \newtheorem{theorem}{Theorem}
\newtheorem{Definition}{Definition}

\newcommand{\bieee}{\begin{IEEEeqnarray}{rCl}}
\newcommand{\eieee}{\end{IEEEeqnarray}}

\SetKwInput{KwInput}{Input}                
\SetKwInput{KwOutput}{Output}              

\begin{document}

\title{On Distribution-Preserving Mitigation Strategies for Communication under Cognitive Adversaries
}

\author{\IEEEauthorblockN{Soumita Hazra and J. Harshan }
\IEEEauthorblockA{\textit{Department of Electrical Engineering,} \\
\textit{Indian Institute of Technology Delhi, India} \\
Soumita.Hazra@ee.iitd.ac.in, jharshan@ee.iitd.ac.in }
}

\maketitle
\begin{abstract}
In wireless security, cognitive adversaries are known to inject jamming energy on the victim’s frequency band and monitor the same band for countermeasures thereby trapping the victim. Under the class of cognitive adversaries, we propose a new threat model wherein the adversary, upon executing the jamming attack, measures the long-term statistic of Kullback-Leibler Divergence (KLD) between its observations over each of the network frequencies before and after the jamming attack. To mitigate this adversary, we propose a new cooperative strategy wherein the victim takes the assistance for a helper node in the network to reliably communicate its message to the destination. The underlying idea is to appropriately split their energy and time resources such that their messages are reliably communicated without disturbing the statistical distribution of the samples in the network. We present rigorous analyses on the reliability and the covertness metrics at the destination and the adversary, respectively, and then synthesize tractable algorithms to obtain near-optimal division of resources between the victim and the helper. Finally, we show that the obtained near-optimal division of energy facilitates in deceiving the adversary with a KLD estimator.
\end{abstract}


\section{Introduction and Problem Statement}

We consider a Denial of Service (DoS)\cite{DOS2} threat on a communication link involving a source, namely Alice, which would like to communicate its messages to the destination, namely Bob in the presence of an active adversary, namely Dave. Dave is a cognitive adversary that injects jamming energy on the frequency band of Alice, and also monitors the same band for potential countermeasures. The idea of monitoring the victim's frequency band for countermeasures is to detect off-the-shelf mitigation methods such as frequency hopping \cite{FHSS}\cite{Jamming1}, which is a popular mitigation scheme against DoS threats. In the context of this work, we are interested in a cognitive adversary \cite{cognitive_radio_jamming3, Conference_Fast_forward, Non_coherent} that is not only capable of monitoring the victim's band, but can also monitor various bands in the network \cite{FDCR1, FDCR2, FDCR3}. With such an adversary, the objective of the victim is to evade the jamming attack and reliably communicate its messages to the destination. Before delving into designing mitigation strategies for the victim, it is imperative to model the process used by the adversary to detect countermeasures. Along those lines, we point out that a long-term statistic based strategy at Dave is to gather the observations on each band before and after the attack, and subsequently, use the two sets of observations to \emph{compare} their statistical distributions. From an information-theoretic viewpoint, this task can be achieved by employing a Kullback–Leibler divergence (KLD) estimator \cite{KLD} on the two distributions. Thus, a problem statement under this cognitive adversarial model is to design mitigation schemes that facilitate Alice to reliably communicate to Bob in the presence of Dave that is equipped with a KLD estimator to detect countermeasures. Henceforth, throughout the paper, a countermeasure is said to achieve covertness with respect to a particular detector if it does not get detected by Dave with an overwhelming probability.

Towards solving the above discussed problem, we make the following contributions. We propose a cooperative strategy wherein Alice, which communicates with On-Off Keying (OOK) signalling, takes the assistance of a helper node, namely Charlie, which is already communicating its messages to Bob using Phase-Shift Keying (PSK). A salient feature of this strategy is that upon detecting jamming, Alice switches her communication to Charlie's frequency band using a fraction of her energy so that Charlie listens to her message and uses a fraction of his energy to forward the same to Bob. Meanwhile the two nodes use a shared secret-key to cooperatively pour their residual energies on Alice's band in such a way that the channel statistics at the victim and the helper bands are nearly identical. The manner in which the two users divide their energies between the two bands is captured by a parameter called the energy division factor, $\alpha \in (0, 1)$. We first show that the proposed strategy is successful in deceiving Dave despite using a KLD estimator on the victim's frequency band irrespective of the choice of $\alpha$. However, to analyze the reliability and the covertness of the proposed strategy on Charlie's frequency band, we notice that the error probability associated with jointly decoding Alice's and Charlie's messages at Bob as well as the probability of detecting a countermeasure at Dave are dependent on $\alpha$. Therefore, in order to compute the optimal $\alpha$ that minimizes their sum, we need to characterize the relation between detection probability and $\alpha$ as a function of the  number of observations used in the KLD estimator. However, given that the frame lengths of the packets are typically short, quantifying the performance of KLD estimator analytically is an intractable task. To circumvent this problem, we propose a stronger countermeasure detector at Dave that is based on comparing the short-term statistic of instantaneous energy on the helper's band before and after the attack. Through this detector, we present rigorous analyses on the error probability at Bob and the detection probability at Dave, and subsequently, propose a near-optimal energy division factor that minimizes their sum. Finally, when using the near-optimal energy values, we also apply KLD based detector at the adversary to show that the estimates are close to zero Importantly, we also show that Alice and Charlie also manage to reliably communicate their messages to the destination, thereby achieving both reliability and covertness. 

The main novelty of this work is the threat model involving an adversary that executes jamming on the victim's frequency and monitor all the network frequencies using KLD estimates and instantaneous energy detector.

\section{Threat Model}
\label{SMPS}
We consider a crowded wireless network wherein all the uplink frequencies  assigned to a destination are allocated to the users of the network. One such instantiation of the crowded network consists of two nodes, namely Alice
and Charlie, that communicate with Bob on two different frequency bands. Alice transmits her
information using OOK over the $f_{AB}$ band, whereas Charlie transmits his information using $M$-ary PSK over the $f_{CB}$ band. Furthermore, Charlie is equipped with a full-duplex radio \cite{FDR2}  with the capability to transmit and receive simultaneously on $f_{CB}$. We also consider an adversary, namely Dave that injects jamming symbols on $f_{AB}$ thereby executing a DoS attack against Alice. In particular, Dave has the following capabilities: (i) In addition to injecting jamming symbols, he is equipped with a full-duplex radio to continuously monitor the statistical distribution of the transmitted symbols of Alice on $f_{AB}$ using a KLD estimator. (ii) He can tune into any uplink frequency and monitor the statistical distribution of its symbols using a KLD estimator. (iii) Furthermore, he has complete knowledge of the constellations used by different users in the network, and therefore, he can also monitor the instantaneous energy of the transmitted symbols on each band. To mitigate this threat, we propose a cooperative strategy involving Alice and Charlie.

\section{Rate-Half Mitigation Strategy} 
\label{RHS}

\begin{figure}[ht!]
\begin{center}
\includegraphics[scale = 0.3]{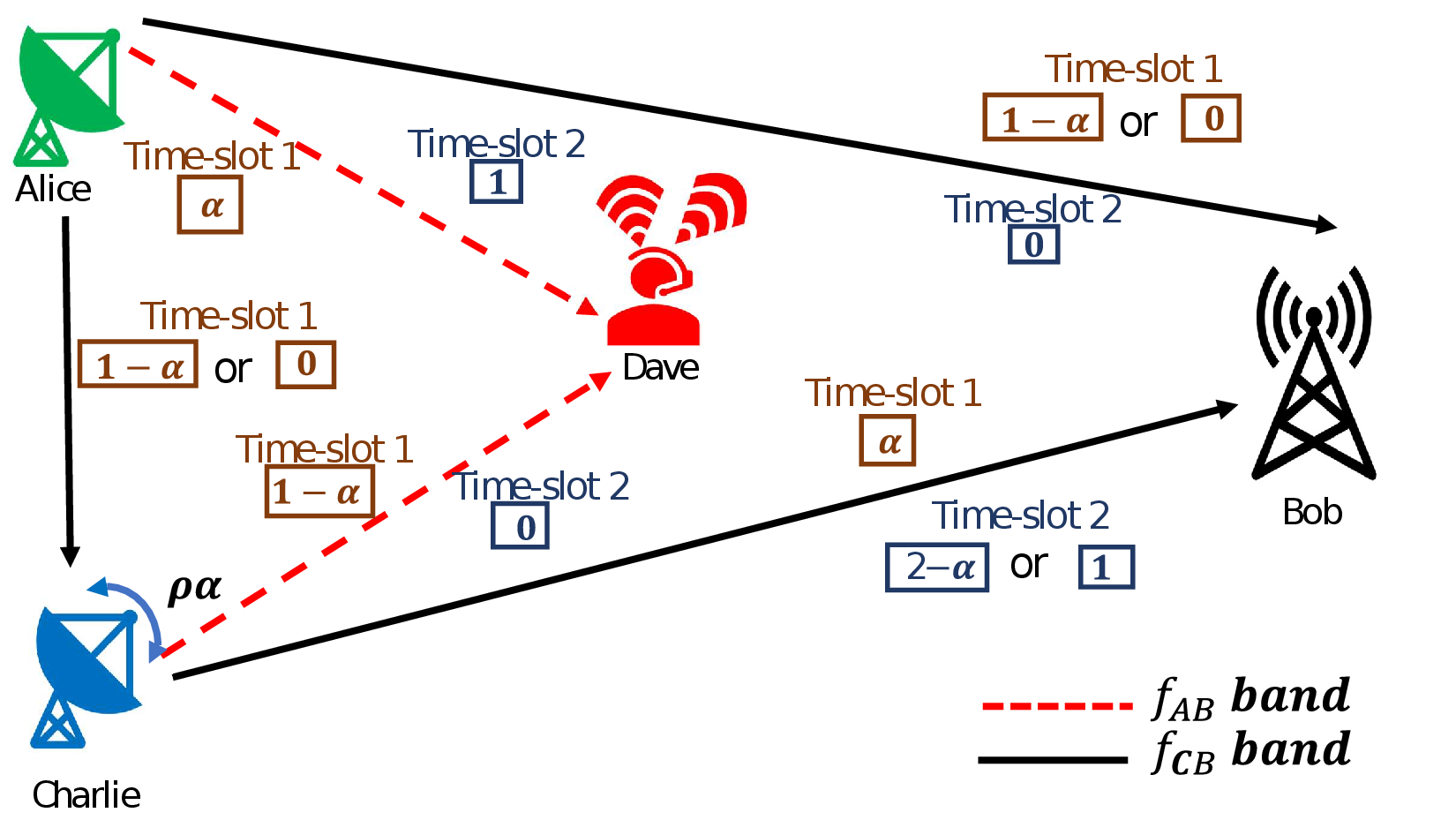}
\end{center}
\vspace{-0.1cm}
\caption{Depiction of the energy levels on both the time slots of the proposed Rate-Half Strategy to mitigate a cognitive adversary.}
\label{Problem}
\end{figure}

In this strategy, Alice and Charlie cooperatively transmit on both $f_{AB}$ and $f_{CB}$ so as to ensure the following two objectives: (i) their information symbols are reliably communicated to Bob on the $f_{CB}$ band, and (ii) their strategy is not detected by Dave despite monitoring the statistical distribution on both $f_{AB}$ and $f_{CB}$ bands. The proposed scheme is divided into two time-slots, as shown in Fig. \ref{Problem}, wherein both Alice and Charlie send one information symbol each in a manner that forbids Dave from detecting this countermeasure with high probability. Since the total number of information symbols sent is half the total number of symbols that would have been sent in the case of no countermeasures, this scheme is termed the Rate-Half strategy. First, we explain the strategy on $f_{CB}$, and then explain the strategy on $f_{AB}$.

\subsection{Strategy on $f_{CB}$ Band}

We ask Alice to transmit her OOK symbols on $f_{CB}$ using a fraction of her energy. Since she is asked to switch to $f_{CB}$ as a reactive measure against jamming, facilitating coherent communication for Alice would result in additional communication-overhead for pilots. As a result, the OOK symbols of Alice can only be decoded in a non-coherent manner. We ask Charlie to continue to communicate his symbols using PSK. Naturally, since Charlie is the incumbent user of $f_{CB}$, we assume that he sends phasor-based pilots that are known only to Bob at regular intervals, and therefore, the PSK symbols can be decoded using a coherent decoder. Our proposed strategy on $f_{CB}$ is divided into two time-slots. 

In the first time-slot, Alice transmits her OOK symbol from the set $\{0, \sqrt{1 - \alpha}\}$, for some $\alpha \in (0, 1)$, which is a design parameter under consideration. If Charlie remains silent in the first time-slot, then Dave would detect a low-energy symbol especially when bit-$0$ is sent by Alice. To circumvent this problem, Charlie also transmits a dummy PSK symbol (already known to Bob), denoted by $\sqrt{\alpha}z_{d}$ in the first time-slot. As a consequence, the received baseband symbol at Bob in the first time-slot is of the form
\begin{equation}\label{eq1:system model}
y_{B1}=\sqrt{1-\alpha}h_{AB}x+\sqrt{\alpha}h_{CB}z_d+n_{B1},
\end{equation}
where $h_{AB}\in\mathcal{CN}(0, 1)$ is the channel between Alice and Bob, $x\in \{0, 1\}$
denotes Alice's bits, $h_{CB}\in\mathcal{CN}(0, 1)$ is the channel between Charlie and Bob, $z_d = e^{-\frac{i 2 \pi j}{M}}, \mbox{for~} j \in \{0,1, \ldots, M-1\}$, denotes the dummy
$M$-PSK symbols transmitted by Charlie and $n_{B1}\in\mathcal{CN}(0, N)$ is the additive white Gaussian noise (AWGN) at Bob in
time-slot 1. Due to a full-duplex radio, the received baseband symbol by Charlie at time-slot 1 is
\begin{equation}\label{eq2:system model}
y_{C1}=\sqrt{1-\alpha}h_{AC}x+ h_{CC}+n_{C1},
\end{equation}
where $h_{AC}\in\mathcal{CN}(0, \sigma^{2}_{AC})$ is the channel between Alice and Charlie with the variance
$\sigma^{2}_{AC}$, $n_{C1}\in\mathcal{CN}(0, N)$ is the AWGN at
Charlie in time-slot 1, $h_{CC} \in \mathcal{CN} (0,\alpha \rho)$ is the loop interference (LI) channel at Charlie, $\rho \in (0, 1)$ is the LI cancellation parameter. 

In time-slot 2, Charlie transmits his PSK symbol, while Alice remains silent. Since Alice is the victim node, it is vital to ensure that Alice’s bits are transmitted with utmost reliability to Bob. Thus, we assume that Charlie uses \eqref{eq2:system model} to recover ${\hat{x}}$, which denotes the decoded bit by Charlie, and then incorporates this bit into his transmitted PSK symbol according to the following rules. When ${\hat{x}}=1$, Charlie will transmit his $M$-PSK symbol without any modification. As a result, the received symbol at Bob in time-slot 2 is given as
\begin{equation}\label{eq3:system model}
y_{B2}=h_{CB}z + n_{B2},
\end{equation}
where $h_{CB}\in\mathcal{CN}(0, 1)$ is the channel between Charlie and Bob, $n_{B2}\in\mathcal{CN}(0, N)$ is the AWGN at Charlie in time-slot 2, $z \in e^{-\frac{i 2 \pi j}{M}}, \mbox{for~} j \in \{0,1, \ldots, M-1\}$, denotes the $M$-PSK symbol transmitted by Charlie. However, if ${\hat{x}}=0$, Charlie will transmit a scaled and rotated version of its $M$-PSK symbol, i.e., with a phase shift of $\frac{\pi}{M}$ and a scale factor of $\sqrt{2-\alpha}$. The corresponding received symbol at Bob is of the form
\begin{equation}\label{eq4:system model}
y_{B2}=\sqrt{2-\alpha}h_{CB}z e^{
\frac{\iota\pi}{M}}+n_{B2}.
\end{equation}
It is worth noting that in time-slot 2, Alice's bit is embedded in the form of a difference in energy level as well as phase-shift from the regular $M$-PSK symbols, thus leading to reliable decoding at Bob's end. Overall, while Charlie's information symbol is communicated in time-slot 2, Alice's information is communicated using both time-slot 1 and time-slot 2. 

\subsection{Strategy on $f_{AB}$ Band}

To tackle the proposed threat model, it is important to maintain OOK symbols on all the time-slots over $f_{AB}$. Therefore, in time-slot 1, we propose Alice and Charlie to cooperatively pour appropriate energy on $f_{AB}$ based on a pseudo-random sequence that is generated using a shared secret-key. When the bit of the pseudo-random sequence is 1, Alice and Charlie, respectively transmit $\sqrt{\alpha}$ and $\sqrt{1 - \alpha}$, thereby resulting in the received symbol $\sqrt{\alpha}h_{AD}+\sqrt{1-\alpha}h_{CD}+n_{D1}$ at Dave, where $h_{AD}\in\mathcal{CN}(0, 1)$, $h_{CD}\in\mathcal{CN}(0, 1)$ are the channels between Alice and Dave, and Charlie and Dave, respectively, $n_{D1}\in\mathcal{CN}(0, N)$ is the AWGN at Dave. On the other hand, if the bit of the pseudo-random sequence is 0, both Alice and Charlie remain silent, thereby resulting in the received symbol of the form $n_{D1}$.

For time-slot 2, we propose that Alice transmits a dummy OOK symbol, denoted by $x_{d} \in \{0, 1\}$, while Charlie keeps silent. The corresponding received symbol at Dave is of the form $h_{AD}x_{d} + n_{D2},$ where $n_{D2}$ is the AWGN at Dave in time-slot 2. Note that the received symbols discussed above are obtained after the removing the LI on $f_{AB}$ at Dave. The following proposition shows that the average energies measured per user and per band are unchanged. 

\begin{Proposition}\label{Energy}
For $\alpha \in (0,1)$, the average energy transmitted over $f_{AB}$ and $f_{CB}$ during the two time-slots are $0.5$ and $1$, respectively. Furthermore, the average
energy that Alice and Charlie contribute over the two time-slots are $0.5$ and $1$, respectively.
\end{Proposition}

\section{Error Analysis at Bob}

Given that Alice's symbols are embedded in both time-slot 1 and time-slot 2, Bob performs joint decoding of the symbols received during the two time-slots on $f_{CB}$. Due to the knowledge of $h_{CB}$ and the dummy $M$-PSK symbol $z_d$, the component $\sqrt{\alpha}h_{CB}z_{d}$ is removed from $y_{B1}$ before the decoding process. The resultant symbol after removing $\sqrt{\alpha}h_{CB}z_d$ is denoted by $\Tilde{y}_{B1} = y_{B1} - \sqrt{\alpha}h_{CB}z_{d}$. Finally, using $\Tilde{y}_{B1}$ and $y_{B2}$, Bob can perform the Joint Maximum A Posteriori (JMAP) decoder given by
\begin{IEEEeqnarray}{rcl}\label{1a}
\hat{a},\hat{b}=\arg \mathop {\max }\limits_{a,b}f\left(\Tilde{y}_{B1}, y_{B2} \left| x \right.=a, z=e^{-\frac{i 2 \pi b}{M}}, h_{CB} \right),
\end{IEEEeqnarray}
where $f\left(\Tilde{y}_{B1}, y_{B2} \left| x \right.=a, z=e^{-\frac{i 2 \pi b}{M}}, h_{CB}\right)$ is the conditional probability density function (CPDF) of $\Tilde{y}_{B1}, y_{B2}$ given $x$, $z$ and $h_{CB}$, and $a \in \{0,1\}$ and $b\in \{0, 1,..., M-1\}$ represent the search space for the joint decoder. The CPDF given in (\ref{1a}) can be written as a combination of Gaussian functions scaled by crossover probabilities introduced at Charlie. However, it is well known that the intricacies in handling Gaussian mixtures makes it challenging to compute the overall error probability of the JMAP decoder given in \eqref{1a}. To circumvent the problem posed by Gaussian mixtures, we propose an approximate JMAP decoder by excluding the terms associated with the cross-over probabilities in the CPDF. Formally, the proposed approximate JMAP decoder, which we refer to as Rate-Half Joint Dominant Decoder (RHJDD), is given by,
\begin{IEEEeqnarray*}{rcl}\label{7a}
\hat{a},\hat{b}=\arg \mathop {\max }\limits_{a,b}f_{JD}\left(\Tilde{y}_{B1}, y_{B2} \left| x \right.=a, z=e^{-\frac{i 2 \pi b}{M}}, h_{CB} \right)
\end{IEEEeqnarray*}
wherein $f_{JD}(\cdot, \cdot)$ denotes the term when the mixture terms associated with error events at Charlie are neglected from $f(\cdot, \cdot)$. Using union bounds on the pair-wise error events, and then averaging the error probability of the RHJDD over several realizations of $h_{CB}$, the following theorem can be stated. 

\begin{theorem}
\label{thm1}
When using RHJDD, an upper bound on the average probability of decoding error at Bob, denoted by $P^{avg}_{UE}$, is given by
\begin{IEEEeqnarray}{rrrcl} \label{P_Eoriginal}
P^{avg}_{UE} =  P_{11} P_{1AVG}+P_{10} P_{1CAVG}+P_{00} P_{2AVG}+ \nonumber \\
P_{01} P_{2CAVG}+P_{11} P_{3AVG}+P_{10} P_{3C},
\end{IEEEeqnarray}
where $P_{mn}$, for $m, n \in \{0, 1\}$, is the probability that Charlie decodes Alice's bit-$m$ as bit-$n$. The other terms are given at the top of next page wherein the underlying parameters are functions of $\alpha$, $N$ and $M$.

\begin{figure*}
\begin{small}
\begin{IEEEeqnarray}{rcl} 
    P_{1AVG}&=&\Bigg[\sum_{i=1}^3 k_{i} exp\left(t_{i}\varphi\right)\sqrt{\frac{\beta_i}{\gamma_i}}\mathcal{K}_1\left(\sqrt{\beta_i\gamma_i}\right)\Bigg] - \Bigg[\sum_{i=1}^{3} k_{i} \sqrt{\frac{\beta_{i}}{\lambda_{i}}}\mathcal{K}_{1}\left(\sqrt{\beta_{i}\lambda_{i}}\right)  exp(\Psi_i)\Bigg], \nonumber
\end{IEEEeqnarray}
\vspace{-3mm}
\begin{IEEEeqnarray}{rcl} 
   P_{1CAVG}&=&1-\Bigg[\sum_{i=1}^3 k_{i} exp\left(-t_{i}\varphi\right)\sqrt{\frac{\beta_i}{\gamma_i}}\mathcal{K}_1\left(\sqrt{\beta_i\gamma_i}\right)\Bigg] - \frac{exp(-A_{b})}{{A_{c}-A_{a}+1}}+\Bigg[\sum_{i=1}^{3} k_{i}\sqrt{\frac{\beta_{i}}{\eta_{i}}}\mathcal{K}_{1}\left(\sqrt{\beta_{i}\eta_{i}}\right)  exp(\Delta_i)\Bigg], \nonumber
\end{IEEEeqnarray}
\vspace{-3mm}
   \begin{IEEEeqnarray}{rcl} 
   P_{2AVG}&=&\Bigg[\sum_{i=1}^3 k_{i} exp\left(-t_{i}\Phi\right)\sqrt{\frac{\mu_i}{\Lambda_i}}\mathcal{K}_1\left(\sqrt{\mu_i\Lambda_i}\right)\Bigg] + \Bigg[\sum_{i=1}^{3} k_{i}\sqrt{\frac{\mu_{i}}{\zeta_{i}}}+
\mathcal{K}_{1}\left(\sqrt{\mu_{i}\zeta_{i}}\right) exp(\varrho_i) \Bigg],\nonumber
\end{IEEEeqnarray}
\vspace{-3mm}
\begin{IEEEeqnarray}{rcl} 
P_{2CAVG}&=&1-\Bigg[\sum_{i=1}^3 k_{i} exp\left(t_{i}\Phi\right)\sqrt{\frac{\mu_i}{\Lambda_i}}\mathcal{K}_1\left(\sqrt{\mu_i\Lambda_i}\right)\Bigg] + \Bigg[\sum_{i=1}^{3} k_{i}\sqrt{\frac{\mu_{i}}{\Omega_{i}}}\mathcal{K}_{1}\left(\sqrt{\mu_{i}\Omega_{i}}\right)
exp(\Upsilon_i)\Bigg]. \nonumber
\end{IEEEeqnarray} 
\vspace{-3mm}  
  \begin{IEEEeqnarray}{rcl}
P_{3AVG}=\sum_{i=1}^3 k_{i}\Big(\frac{t_{i}}{N_{0b}}+1\Big)^{-1},\hspace{2mm} P_{3C}&=&\frac{1}{2}\nonumber.
\end{IEEEeqnarray}
\end{small}
\hrule
\end{figure*}
\end{theorem}

\section{Covertness Analysis}
\label{PDs}

We discuss the accuracy with which the proposed countermeasure can be detected at Dave. Although Dave does not know the frequency band of Charlie, we restrict our study to only $f_{AB}$ and $f_{CB}$ since other bands are implicitly unaltered. Recall that Dave has the ability to monitor the statistical distributions on $f_{AB}$ and $f_{CB}$ by using a KLD estimator based detector. With respect to covertness on $f_{AB}$, the following theorem can be proved. 

\begin{Proposition}\label{fab}
The statistical distribution of the symbols on $f_{AB}$ after implementing the Rate-Half strategy is identical to that before the countermeasure. 
\end{Proposition}

For covertness on $f_{CB}$, while a KLD estimator can be used to compare the statistical distributions before and after the attack, characterizing its performance is intractable with finite number of samples. Therefore, we propose a short-term metric based detector, wherein Dave monitors the instantaneous energy level on $f_{CB}$. To achieve this, we recall that Charlie broadcasts pilot symbols to Bob at regular intervals by using a pre-shared phasor symbols that is unknown to Dave. Although Dave would not be able to estimate the channel between Charlie and itself, we make a worst-case assumption in the benefit of Dave that he can estimate the magnitude of the channel. As a result, in the case of no countermeasure, the received symbol at Dave on $f_{CB}$ is of the form
\begin{IEEEeqnarray}{rcl}\label{no_countermeasure}
y_D=h_{CD}y+n_{D},
\end{IEEEeqnarray}
where $h_{CD}\in \mathcal{CN} (0,1)$ is the channel between Charlie and Dave, $y \in e^{-\frac{i 2 \pi j}{M}}, \mbox{for~} j \in \{0,1, \ldots, M-1\}$ denotes the $M$-PSK symbol and $n_{D} \in \mathcal{CN}(0, N)$ is the AWGN at Dave. On dividing (\ref{no_countermeasure}) by $|h_{CD}|$, we obtain
\begin{IEEEeqnarray}{rcl}\label{no_countermeasure1}
y_D^{'}=ye^{\angle h_{CD}}+n_{D}^{'},
\end{IEEEeqnarray}
where $y_D^{'}=\frac{y_D}{|h_{CD}|}$ and $n_{D}^{'} \in \mathcal{CN}\left(0,\frac{N}{|h_{CD}|^2}\right)$ is the effective AWGN at Dave. In the absence of AWGN, the energy of the received symbol $|y_D^{'}|^2$ lies on the circumference of a unit circle. However, due to the presence of AWGN, the received energy may lie around the unit circle with a majority of energy lying in between $1-\delta$ and $1+\delta$, for some $\delta > 0$. Towards detecting any possible countermeasure, Dave can use this behaviour to expect $|y_D^{'}|^{2}$ within $1-\delta$ and $1+\delta$, and subsequently, raise a detection event if $|y_D^{'}|^{2} > (1+\delta)$, or $|y_D^{'}|^{2} < (1-\delta)$. Naturally, in the event of no countermeasure, the optimal value of the allowed energy deviation is the value of $\delta$ for which the probability of false alarm is bounded by a small number of Dave's choice. We formally define probability of false alarm in \textit{Definition} \ref{PFA_d} given below.

\begin{Definition} \label{PFA_d}
Under the hypothesis that no countermeasure is implemented, for a given $\delta > 0$ and $|h_{CB}|$, the probability of false alarm, denoted by $P_{FA}$, is given by
\begin{IEEEeqnarray}{rcl}\label{n_C4}
P_{FA} & = & \Pr\{|y_D^{'}| \geq\sqrt{1+\delta}\}+\Pr\{|y_D^{'}| \leq\sqrt{1-\delta}\}.
\end{IEEEeqnarray}
\end{Definition}

We notice that deriving the CPDF on $|y_D^{'}|^{2}$ is a challenging task. As a result, we take the approach of upper bounding $P_{FA}$ by using some upper bounds and lower bounds on $|y_D^{'}|^{2}$. In particular, we use the upper bound $|y_D^{'}| \leq  |n_{D}^{'{}}|+1$ and the lower bound $|y_D^{'}| \geq ||n_{D}^{'{}}|-1|$ in the first and the second term of \eqref{n_C4}, respectively, to obtain an upper bound on $P_{FA}$ as 
\begin{IEEEeqnarray*}{rcl}\label{n_C5}
P_{FA} \leq \Pr\{|n_{D}^{'{}}|+1\geq\sqrt{1+\delta}\}
+\Pr\{|1-|n_{D}^{'{}}|| \leq\sqrt{1-\delta}\}.
\end{IEEEeqnarray*}
Using the above expression, we obtain the following result.

\begin{Proposition} For a given $\delta$, an upper bound on the average probability of false alarm, denoted by $P^{avg}_{UF}$,  is given by

\begin{small}
\begin{IEEEeqnarray}{rcl}
P^{avg}_{UF} &=& \left(1+\frac{(\sqrt{1+\delta}-1)^2}{N}\right)^{-1} 
+ \left(1+\frac{(-\sqrt{1-\delta}+1)^2}{N}\right)^{-1} \nonumber \\
&-&\left(1+\frac{(\sqrt{1-\delta}+1)^2}{N}\right)^{-1}.\nonumber
\end{IEEEeqnarray}
\end{small}
\end{Proposition}

\begin{figure*}
\begin{equation*}
P_{D_{11AVG}} =\left(1+\frac{J_1}{N_{1b}}\right)^{-1}
+ \left(1+\frac{J_2}{N_{1b}}\right)^{-1}
- \left(1+\frac{J_3}{N_{1b}}\right)^{-1}, P_{D_{10AVG}} =\left(1+\frac{J_1}{N_{0b}}\right)^{-1}
+ \left(1+\frac{J_2}{N_{0b}}\right)^{-1}
- \left(1+\frac{J_3}{N_{0b}}\right)^{-1}
\end{equation*}
\begin{equation*}
P_{D_{20AVG}} = \left(1+\frac{J_4}{N_{0b}}\right)^{-1}
 + \left(1+\frac{J_5}{N_{0b}}\right)^{-1}
- \left(1+\frac{J_6}{N_{0b}}\right)^{-1}, P_{D_{21AVG}}=P^{avg}_{UF}.
\end{equation*}
\hrule
\end{figure*}

Using the above proposition, Dave can choose $\delta > 0$ such that $P^{avg}_{UF}$ is bounded by a small number of his choice. In the rest of this section, we discuss the probability with which the proposed countermeasure would be detected by Dave for a given $\delta$. With $y_{D1}$ and $y_{D2}$ denoting the symbols received at Dave in time-slot 1 and time-slot 2 on the frequency band $f_{CB}$, we have  

\begin{small}
\begin{equation}
	\label{eq:rx_dave_with_count_1}
y_{D1}^{ } = \left\{ \begin{array}{cccccccccc}
		\sqrt{1-\alpha}h_{AD} +\sqrt{\alpha}h_{CD}z_{d}+n_{D1}, & \mbox{ if } x=1;\\
		\sqrt{\alpha} h_{CD} z_d + n_{D1}, & \mbox{ if } x=0;\\
	\end{array}
	\right.
\end{equation}
\end{small}
\begin{equation}
	\label{eq:rx_dave_with_count_2}
y_{D2}^{ } = \left\{ \begin{array}{cccccccccc}
		h_{CD}z +n_{D2}, & \mbox{ if } \hat{x}=1;\\
		\sqrt {2-\alpha }h_{CD}e^{\iota\frac {\pi }{M}}z + n_{D2}, & \mbox{ if } \hat{x}=0;\\
	\end{array}
	\right.
\end{equation}
where $h_{AD}, h_{CD} \in \mathcal{CN}(0, 1)$ are the channels between Alice and Dave, and Charlie and Dave, respectively. Similarly, $n_{D1}, n_{D2}\in \mathcal{CN}(0, N)$ are the AWGN at Dave in time-slot 1 and time-slot 2, respectively. The other variables in \eqref{eq:rx_dave_with_count_1} and \eqref{eq:rx_dave_with_count_2} follow from the proposed countermeasure. As per the detection strategy, Dave uses $|h_{CD}|$ to compute $y_{D1}^{'} = y_{D1}/|h_{CD}|$ and $y_{D2}^{'} = y_{D2}/|h_{CD}|$, and then verifies whether $|y_{D1}^{'}|^2$ and $|y_{D2}^{'}|^2$ lie outside the concentric circles with radii $(1-\delta)$ and $(1+\delta)$. The following definition captures the probability of detection.

\begin{Definition}
Under the hypothesis that the proposed countermeasure is implemented, the probability of detection is the probability that either $|y_{D1}^{'}|^2$ or $|y_{D2}^{'}|^2$ lie outside the concentric circles with radii $(1-\delta)$ and $(1+\delta)$.
\end{Definition}

Using the above definition, the following theorem provides an upper bound on the average probability detection.

\begin{theorem}\label{PDavg}
When $1-\delta < \alpha$, an upper bound on the average probability of detection is given by
\begin{IEEEeqnarray*}{rcl}\label{PD_avg}
P^{avg}_{UD} = \frac{1}{2}[P_{D_{10AVG}} + P_{D_{11AVG}} + (P_{00}+P_{10})P_{D_{20AVG}}\\
 + (P_{11}+P_{01})P_{D_{21AVG}}]\nonumber,
\end{IEEEeqnarray*}
where the individual terms are listed at the top of this page such that $J_{1}=(\sqrt{1+\delta}-\sqrt{\alpha})^2$, $J_{2}=(-\sqrt{1-\delta}+\sqrt{\alpha})^2$, $J_{3}=(\sqrt{1-\delta}+\sqrt{\alpha})^2$, $J_{4}=(\sqrt{1+\delta}-\sqrt{2-\alpha})^2$, $J_{5}=(-\sqrt{1-\delta}+\sqrt{2-\alpha})^2$, and $J_{6}=(\sqrt{1-\delta}+\sqrt{2-\alpha})^2$, where $N_{0b} = N$ and $N_{1b} = N + 1-\alpha$. 
\end{theorem}

\section{Near-Optimal Energy division factor} \label{Algo}

\begin{figure}[ht!]
\begin{center}
\includegraphics[width=8cm,height=4cm]{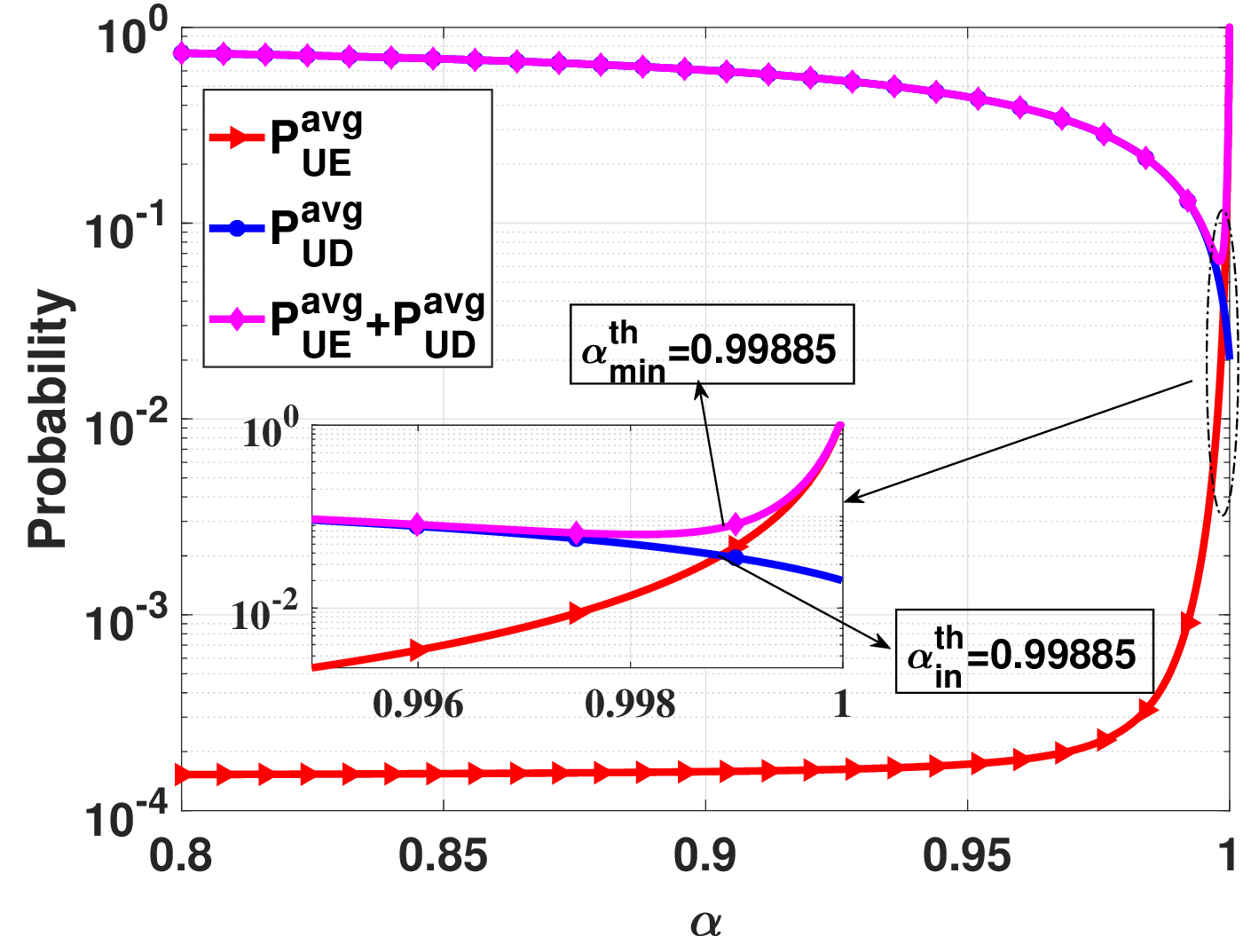}
\end{center}
\vspace{-0.3cm}
\caption{Figure shows that the intersection point between $P_{UE}^{avg}$ and $P_{UD}^{avg}$ is close to the minima of their sum, when using the parameters $\delta=0.495$, $P^{avg}_{UF}=10^{-2}$ at signal-to-noise-ratio of $35$ dB.}
\label{Intersection}
\end{figure}

\begin{figure}[ht!]
\subfloat{\includegraphics[width=4.2cm,height=2.3cm]{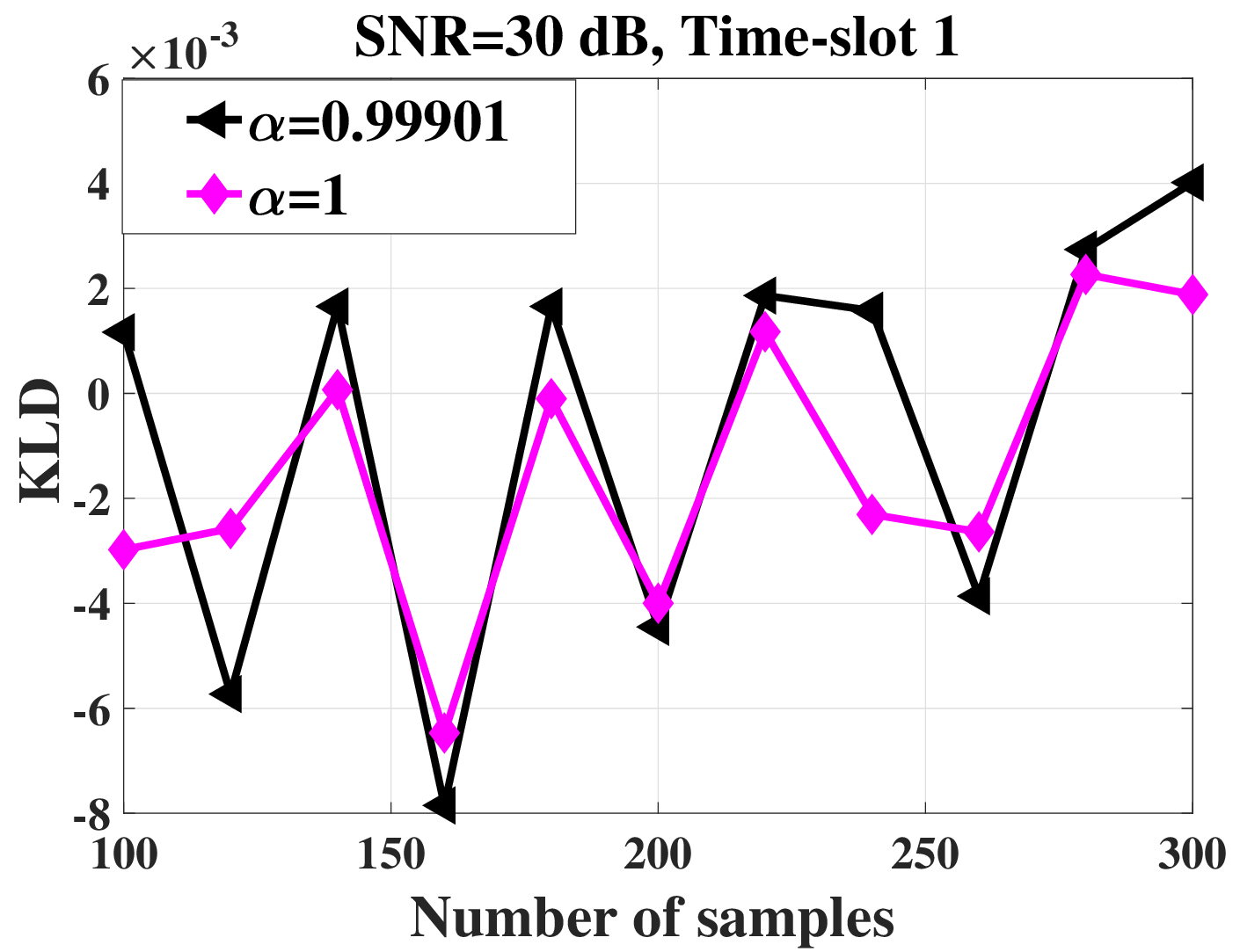}}
\hspace{4mm}
\subfloat{\includegraphics[width=4.2cm,height=2.3cm]{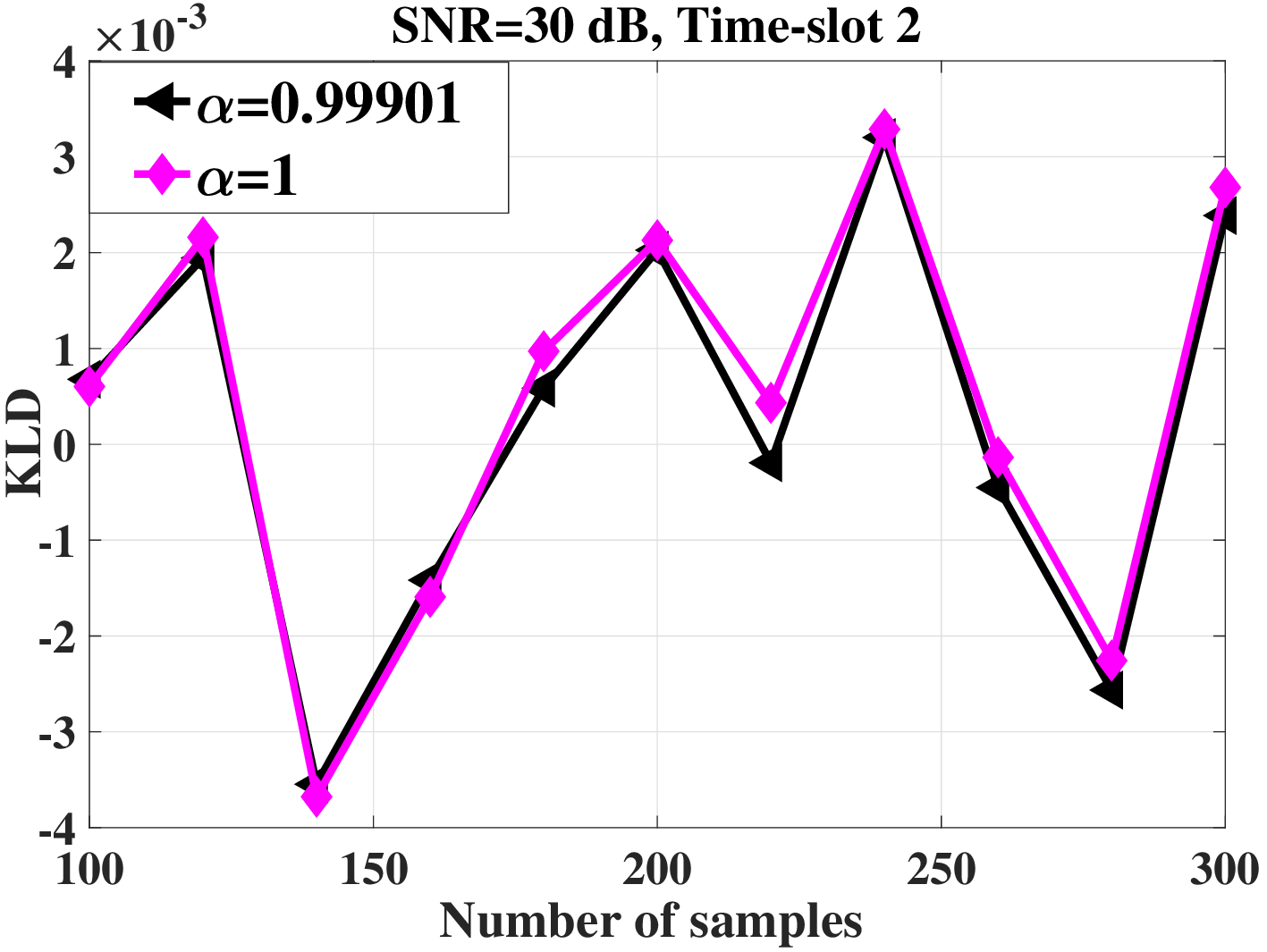}}

\subfloat{\includegraphics[width=4.2cm,height=2.3cm]{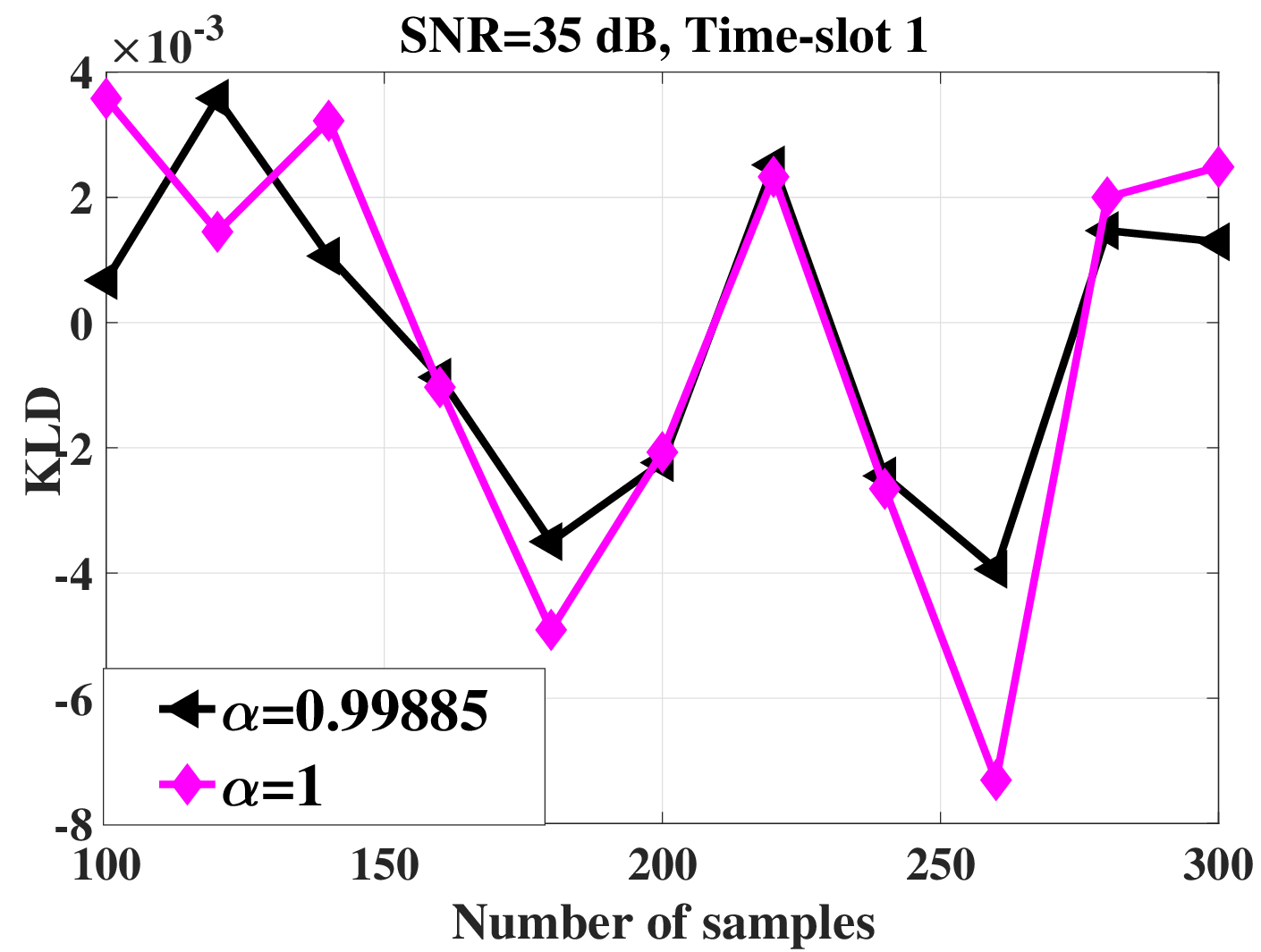}}
\hspace{4mm}
\subfloat{\includegraphics[width=4.2cm,height=2.3cm]{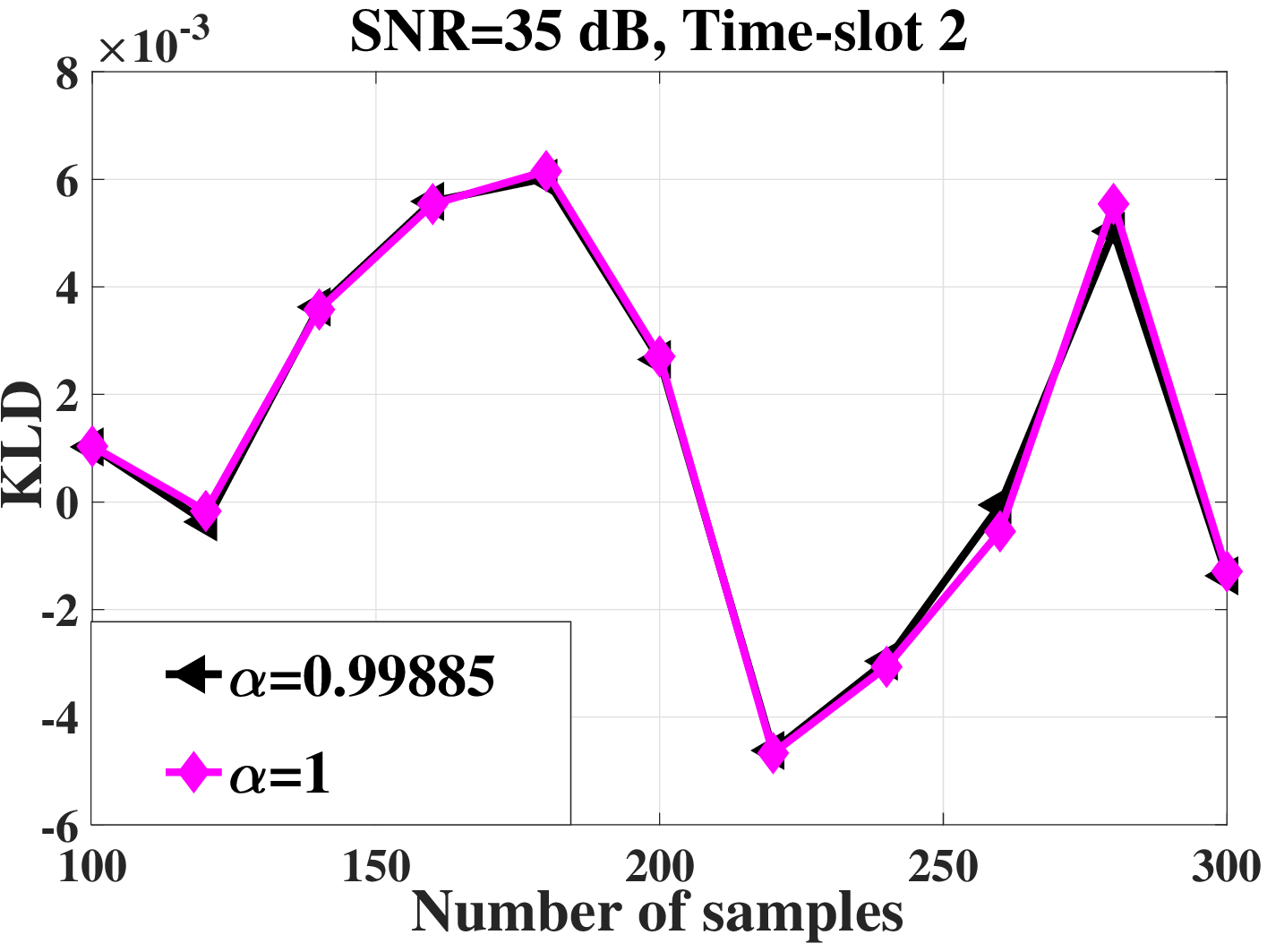}}
\caption{Average KLD metric when computed for both time-slots at signal-to-noise-ratio of $30, 35$ dB, for the parameters used in Fig. \ref{Intersection}.}
\label{KLD}
\end{figure}

In this section, we identify the behaviour of $P^{avg}_{UE}$ and $P^{avg}_{UD}$ with respect to $\alpha \in (0, 1)$, and then propose a method to compute an appropriate value of $\alpha$ for implementation. Based on the proposed strategy on $f_{AB}$ and $f_{CB}$, it is clear that as $\alpha \rightarrow 0$, detection probability of the instantaneous energy detector is high, whereas the average probability of error at Bob is negligible. On the other hand, as $\alpha \rightarrow 1$, detection probability of the instantaneous energy detector is low, whereas the average probability of error at Bob is high. To communicate both reliably and covertly, it would be interesting to minimize $P^{avg}_{UE} +P^{avg}_{UD}$ over $\alpha \in (0, 1)$ for a given bound on $P^{avg}_{UF}$. However, given the complex nature of the expression on $P^{avg}_{UE} +P^{avg}_{UD}$, we notice that analytically solving the minima of the objective function is intractable. We also notice through several simulation results (as exemplified in Fig. \ref{Intersection}) that solving for the intersection between $P^{avg}_{UE}$ and $P^{avg}_{UD}$ would give us an $\alpha$ close to the minima. Therefore, we propose to solve $P_{UE}^{avg}-P_{UD}^{avg}=0$ subject to $\alpha \in (0, 1)$ by using iterative algorithms such as the Newton-Raphson (NR) method.  
It is interesting to observe from Fig. \ref{Intersection}, that with the choice of $\alpha = 0.99885$, the proposed Rate-Half strategy achieves an error rate of the order of $10^{-2}$ along with the same probability of detection when using the instantaneous energy detector at Dave. We remark that lower values of error- and detection-rates can be achieved when the reliability of Alice to Charlie link improves thereby pushing the point of intersection further close to $1$. Interestingly, when using the KLD estimator for detection with $\alpha = 0.99885$, we show through Fig. \ref{KLD} that the average KLD metric is very close to zero on both the time-slots thereby keeping the statistical distributions of the observations approximately same before and after the attack. 

\hspace{-5mm}

\vspace{-0.8cm}
\section{Discussion}

While long-term statistics based KLD estimator based detector is optimal from an information-theoretic viewpoint, one of the challenges for future research is to characterize its probability of detection with finite number of samples in wireless settings. Success along these lines will help us analyze the Rate-Half strategy and also derive its optimal energy division factor without taking the assistance of instantaneous energy detector, which is based on short-term statistics.

\end{document}